# Possible superconductivity in brain

P. Mikheenko[*]

Department of Physics, University of Oslo,  P.O.Box 1048, Blindern, 0316, Oslo, Norway
*pavlo.mikheenko@fys.uio.no

The unprecedented power of brain suggests that it may process information quantum-mechanically. Since quantum processing is already achieved in *superconducting* quantum computers, it may imply that superconductivity is the basis of quantum computation in brain too. Superconductivity could also be responsible for long-term memory. Following these ideas, the paper reviews the progress in the search for superconductors with high critical temperature and tries to answer the question about the superconductivity in brain. It focuses on recent electrical measurements of brain slices, in which graphene was used as a room-temperature quantum mediator, and argues that these measurements could be regarded as evidence of superconductivity in the neural network of mammalian brains. The estimated critical temperature of superconducting network in the brains is rather high: 2063±114 K. A similar critical temperature was predicted in the Little's model for one-dimensional organic chains linked to certain molecular complexes. A reasonable suggestion is that superconductivity develops in microtubules inside the neurons of brain.

## Introduction

Superconductivity in brain was suggested in 1972 in order to explain long-time memory and following a general argument that if room-temperature superconductivity is existing, it should be in a system with high level of organization [1]. In recent advance, superconductivity proved to be successful in quantum computing [2-4]. Combining these facts with novel idea that brain is quantum computer [5,6], one could deduce that superconductivity might be at work in brain too. A problem with this statement is that superconductivity was not yet proven to exist at room temperature. There is, however, no reason to believe that it cannot exist at this temperature. Recent progress in search for superconductivity in hydrates [7] shows that room-temperature superconductivity is quite realistic [8,9]. Broadly speaking, brain can also be classified as complex, well-organized hydrates-based material. There is, however, substantial difference between brain structures and already analyzed superconducting hydrates. This difference will be discussed in the review and its implication for possible room-temperature superconductivity will be outlined.

## I. Search for superconductors with higher critical temperature

Since its discovery in 1911 [10] (Nobel Prize in physics for the year 1913), superconductivity was considered to be essentially low-temperature phenomenon. The critical temperature ($T_c$) of the first superconducting material, Hg, was about 4 K, and during next 75 years, in spite of intensive search for materials with higher critical temperature, it increased only by 20 K. There were many reasons for such apparently slow progress. As it is seen from Fig. 1, where a sketch of the increase in critical temperature is outlined, first, quite logical, elemental materials were explored. The highest $T_c$ was found in Nb (9.3 K [11], 8.4 K in first measurements [12]). Surprisingly, most conductive materials: Ag, Au and Cu did not show superconductivity.

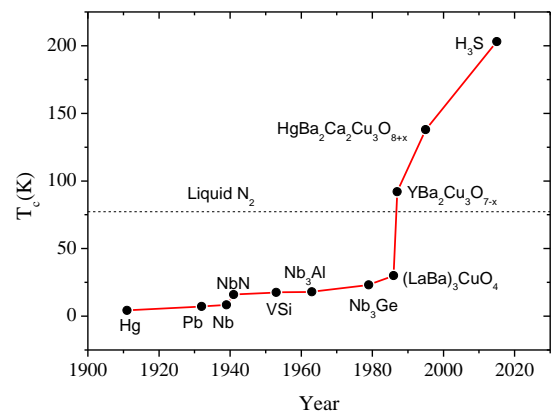

Fig. 1. A sketch of the progress in increasing critical temperature of superconducting materials. The dashed line shows boiling temperature of liquid nitrogen.

At about 1940, the complexity of investigated materials was increased and the efforts were concentrated on binary compounds, mainly on those containing Nb, as this element showed the highest critical temperature. There was immense progress in mastering Nb-based materials. They are up to now most-used compounds in applications of superconductivity. In these materials, $T_c$ increased to 23.2 K in $Nb_3Ge$ [13]. However, due to this focus, one very important binary compound, $MgB_2$, with $T_c$ of 39 K [14], which could form basis of renewal hydrogen economy [15], was lost for about five decades. One could also wonder if superconductivity in binary compound Au-Ag was also lost, following recent claim of nearly room-temperature superconductivity in this material [16,17].

The discovery of high-temperature superconductors seen on Fig. 1 as sharp increase in $T_c$ in 1986 (Nobel Prize in physics to Bednortz and Muller in 1987) that crossed the line of boiling temperature of liquid nitrogen, most abundant gas in Earth's atmosphere, is very important milestone. Finding this whole new class of materials could be considered as a result of increasing complexity, i.e. moving in research from binary compounds to compounds with four elements. A strategy leading to discovery of superconductivity in these materials was, however, different, namely searching for strong polarizability and polaronic effects [18]. This search for specific physical effects that could be precursors of superconductivity is perhaps most important principle able to lead to breakthroughs. Most recent example, featuring record-high $T_c$ in the family of hydrates [7-9] is very good illustration of this principle. In this case, increase of $T_c$ is due to the presence of light atom (hydrogen) and the shift of the mechanism of pairing from acoustic to optical phonons [9]. Getting knowledge about specific physical effects, modern physics is now able to predict superconducting properties of materials before their experimental study [8].

Analyzing mechanisms leading to superconductivity, several pathways to materials with possible room-temperature superconductivity are outlined [19]. For successful research, one should choose a right combination of these paths, perhaps switching from one to another on a way to a new compound. A specific guideline highlighted in this review is reduction of dimensionality of the materials, in which one hopes to find extra high $T_c$.

This suggestion may sound not logical, since it is well known that in lower-dimension materials, *i.e.* one- and two-dimensional (1D, 2D), fluctuations of the order parameter are so strong that they prevent long-range quantum order [20,21], including that responsible for superconductivity. However, lower-dimension systems offer greater flexibility and the superconductivity could be possible if one manages to tackle these fluctuations.

Among the pioneers of research in low-dimensional materials were Kosterlitz and Thouless [22], who proved that superconductivity is possible in two dimensions. In this case, fluctuations (vortices and antivortices) are bound together below $T_c$. In 2016, Kosterlitz and Thouless were awarded the Nobel Prize in physics. It is interesting that earlier mentioned Nobel Prize awarded to Bednortz and Muller was also for materials with reduced dimensionality. After the discovery, it was soon found that superconductivity in this class of materials is quasi-two dimensional. $T_c$ in these materials reached 164 K at high pressure [23], which was much higher than in known that time 3D systems. Using high pressure, however, would allow to have very high $T_c$ in 3D also [8,9], although practicality of high-pressure materials is currently questionable.

Following pathway of reduced dimensionality, one may ask about possible high $T_c$ in quasi-1D systems. This case was, again, analyzed theoretically. As early as in 1964, Little [24] showed that superconductivity in linear chains of organic molecules linked to certain molecular complexes could have $T_c$ of ≈2200 K. It does not mean that one could observe superconducting transition directly, as the system would be destroyed long before reaching $T_c$, but this does not prevent existence of superconductivity at ambient temperature, at which organic molecular chains are stable.

It is important to note that in the model of Little the superconducting pairing is provided by electron-electron interaction, in contrast to more common electron-phonon interaction (Nobel Prize in Physics to Bardeen, Cooper and Schrieffer for the year 1972). In case of Little's model, at least three particular paths: lower dimensionality, specific pairing interaction and increased complexity were followed.

To display such high $T_c$ as in the model of Little, one needs to modify Fig. 1, for example, presenting y-axis in the logarithmic scale, as shown in Fig.2. The highest point in this figure belongs to main experimental result of the review, which appears to be very close to the prediction of Little. The rest of the paper will be focused on the activity related to this result, namely on hypothesized superconductivity in brain.

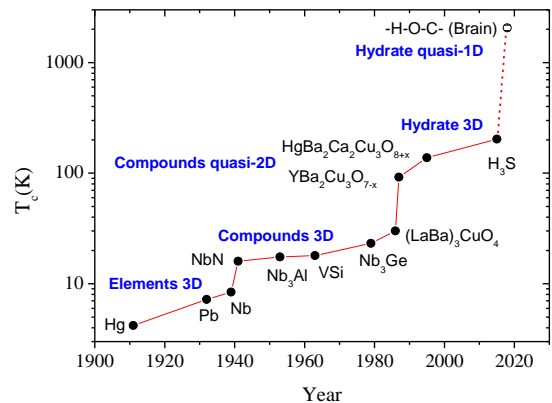

Fig. 2. A logarithmic presentation of the critical temperature growth for milestone superconducting materials. The highest point belongs to main material discussed in this review.

## II. Superconductivity in brain

Increasing complexity, which is important pathway to superconductivity with high $T_c$, inevitably leads to most-organized structures of living organisms: nervous system and brain [1]. While humans are consciously experimenting with superconductivity for about 100 years, nature might be doing this subconsciously for billions of years, perfecting molecular structures from generation to generation and arriving to most efficient coherent structures able to process information in quantum-mechanical way.

Search for superconductivity in organic and biological materials was active for a long time [24-28]. Specific recent examples can be found in [19] and references therein. There are, however, currently no convincing experimental evidences of superconductivity in brain and nervous system. In order to provide such evidence, specific experimental techniques need to be developed. These should allow quantum-mechanical study of strongly fluctuating systems at room temperature. One of such techniques was, in fact, already developed. In [29], it was argued that for electrical transport measurements of room-temperature superconductor, a mediator that behaves quantum-mechanically at room temperature is needed. In application to biological systems, such mediator should also be highly conductive and able to penetrate membranes to provide connections to organic molecular chains. Such a mediator became available only recently. It is graphene, the Nobel Prize winning (2010) material [30]. Graphene can be prepared in aqueous solutions compatible with the natural environment of brain cells.

The graphene-based technique was already applied to a rat's brain slice [29]. Since it is not possible to measure superconducting transition directly due to expected very high $T_c$, the study was focused on current-voltage (IV) characteristics. IV curves of a formaldehyde-fixed brain slice measured at room temperature, indeed, showed similarity with IV curves of a superconductor [29]. Later this activity was extended to pigs' brains. The obtained data could be considered as evidence of room temperature superconductivity in brain. To proceed further, it is necessary to describe this experimental technique in more detail.

### III. Experiments on brain slices

The electrical measurements of the slices of pigs' (*Sus domesticus*) brains were carried out by four-electrode technique designed to connect (via the nano-plates of graphene) conductive channels in the brain. The possible types of connection are schematically shown in Fig. 3 together with photos displaying contact wires (d), investigated brains (e) and a sketch of neural network in a brain slice attached to the contacts (f).

The measurement system features large-area contacts of neural network to thick (0.2-0.5 mm in diameter) current and potential wires shown in Fig 3d. The length of the wires is about 1 cm. Several millions microtubules are connected in parallel to the wires via graphene nanoflakes, as shown schematically in Fig. 3f. The slices of the brains were pressed to the contacts after being wrapped in a plastic film to reduce evaporation of water. The measurements give statistically averaged values of voltage and current in the network.

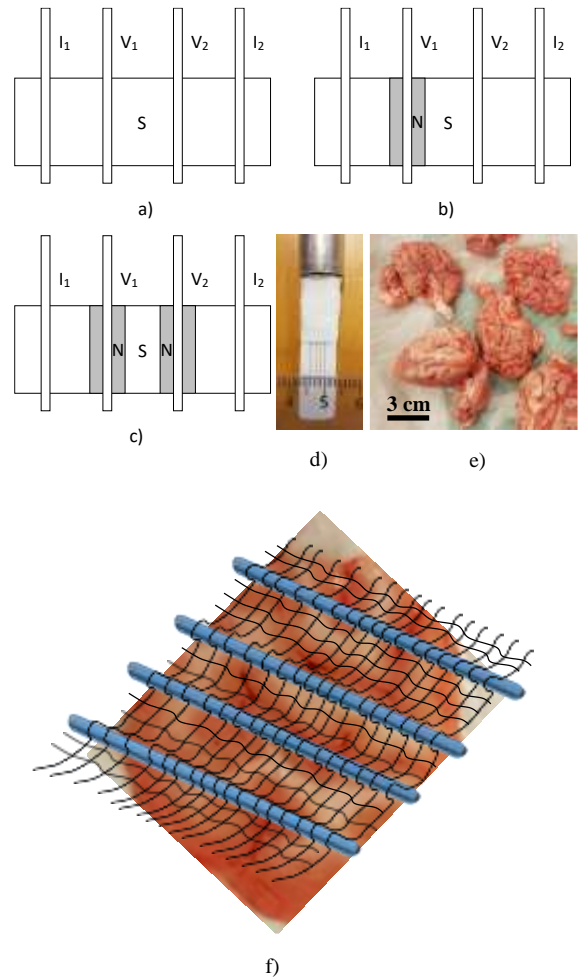

Fig. 3. Schematic representation of the arrangement of current ($I_1$, $I_2$) and potential ($V_1$, $V_2$) electrodes during measurements of brains' slices (horizontal rectangles). Possible suppression of superconductivity (S) by potential electrodes resulting in normal region/s (N) is taken into account. In a), there is no suppression. In b), superconductivity is suppressed close to $V_1$ only, and in c), superconductivity is suppressed close to both potential electrodes. d) A photo of contact wires to which the slices of brain were pressed. e) The pigs' brains used in experiments. f) Schematic picture showing connection of neurons to electrodes.

The central feature of the technique is use of graphene as a room-temperature quantum mediator, which allows transferring information from a quantum object to the classical measurement system. The water suspension of graphene became recently available from commercial suppliers. Graphene is an exciting novel material (the Nobel Prize in physics for the year 2010). It is a highly conductive single-atom thick nano-material with carbon atoms arranged in a honeycomb lattice. In the solution used in these experiments (Sigma-Aldrich, graphene nanoplatelets 1 mg/mL, dispersion in $H_2O$), graphene is present in the form of flakes with a range of in-plane sizes from fractions of micron to a few microns.

A range of current and potential wires made of copper, copper covered with a Pb-Sn alloy and gold were used as electrodes. Altogether, 26 slices of the thickness of about 1 mm of *Sus domesticus'* brains from the samples shown in

Fig. 3e were measured. The brains were fast-delivered (in 3 hours) from a meat-supplying company and measured fresh after being exposed to graphene solution for different periods of time. Some brains were fixed with formaldehyde and used in the experiments later. Control measurements of filter paper soaked in graphene and brain's slices without graphene were also carried out. Complementary measurements of a dry sample of bacterial derivative, to check operation of system, and non-brain (muscle) wet tissues of *Gallus gallus domesticus* were carried out earlier and reported in [29]. No features of superconductor-like behaviour have been observed in control samples. Magneto-optical images obtained in [29] showed diamagnetic response, which could be expected from a sample containing superconducting network.

The carried transport measurements rely on possibility that nano-structures in the brain responsible for hypothesized superconductivity are preserved after the preparation of slices and the samples remain in superconducting state for sufficiently long time to be measured. It is also taken into account that superconductivity could be suppressed in the vicinity of electrodes. Following these suggestions, in Figs.3a-c possible outcomes of measurements are outlined. In the best case (a), connections between one-dimensional channels of the superconducting network and potential electrodes are intact, superconductivity is not suppressed and one would expect to obtain standard current-voltage characteristic of a superconductor with a dissipationless part.

If the superconductivity is suppressed close to one of the potential electrodes (b), a finite resistance should always be present in the measurement. The largest contribution to it may come from the superconductor-normal metal (SN) interface, on which the current flow is strongly suppressed for voltages below $\Delta/e$, where $\Delta$ is the energy gap of the superconductor. At $\Delta/e$, a decrease in resistance should take place, because at this voltage charge carriers are allowed to fill empty states above the energy gap.

In the third case, superconductivity is suppressed close to both electrodes (c). Resistance is, again, finite, but it is expected to decrease at voltage $2\Delta/e$ corresponding to two NS boundaries. In the case of a 1D superconducting network, there is a very large number of connections to both electrodes, and the configurations outlined in Fig. 3 give statistically averaged features. In an experiment, one could expect fluctuating behavior between configurations b) and c), *i.e.* a partial decrease in resistance at both voltages: $\Delta/e$ and $2\Delta/e$. To access pure superconducting state, configuration a) should be realized.

The simplicity of outlined description comes from expected large value of $\Delta$, in which case serial resistance of the normal phase N in Fig. 3 could be neglected. The energy-gap anomalies, if observed, could be a good argument in favor of superconductivity. The value of gap, when determined, would also allow estimating $T_c$.

## IV. Measurements of energy gap and estimation of critical temperature

Current-voltage (IV) characteristics of a formaldehyde-fixed slice of rat's brain of the thickness of 40 microns exposed to graphene solution for 1 minute and recorded at room temperature in two areas of different width are shown in Fig. 4. The set of curves for the larger cross-section (black line, blue points, right axis) is scaled in current by a factor of 600 (refer to the left axis for the curves with smaller cross-section). IV curves for both cross-sections were recorded during an increase and decrease of voltage and showed good reproducibility. A specific feature for IV curves in both cross-sections are large voltage jumps of equal length shown by the horizontal arrows in the main plot and in the inset, where IV curves of the higher-current part of the sample with the larger cross-section are plotted separately for clarity.

A remarkable fact is that in spite of a nearly three order of magnitude difference in current, the voltage jumps are of the same magnitude, about 0.6 V. For the IV curve of a superconductor, such jumps could appear during thermomagnetic instability-driven local suppression of superconductivity. The minimum voltage jump for the instabilities is $2\Delta/e$ [31], which is the energy per unit of charge necessary to cross two NS boundaries. With this interpretation and following similarity of recorded IV curves with those in thin-film conventional superconductors [31], Fig. 4 provides an estimate of the energy gap. Moreover, since there is a direct link between the energy gap and critical temperature, $T_c$ could be estimated too.

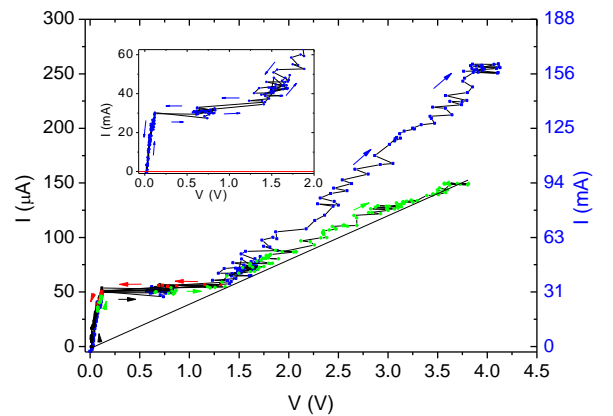

Fig. 4. Current-voltage characteristics of a slice of rat's brain of the thickness of 40 microns recorded at room temperature in two areas of different width. The scale for the larger cross-section (black line, blue points) is on the right axis and for the smaller cross-section on the left one. At the same voltages, the current through the larger cross-section is about 600 times higher than through the smaller cross-section. Curves for both cross-sections are recorded during an increase and decrease of voltage, as shown by the small arrows. In spite of the large difference in current, the two large voltage jumps shown by the horizontal arrows are of the same magnitude in both sets of curves. The inset shows the low-voltage parts of the IV curves for the larger cross-section, illustrating the magnitude of the voltage jumps. A thin line in the main plot is drawn for the guide of eye.

To link Δ and $T_c$, one can use standard formula for phonon-mediated superconductivity [19]:

$$2\Delta(0)/e = 3.53\, k_B T_c, \quad (1)$$

where $k_B$ is the Boltzman constant and $\Delta(0)$ is the energy gap at zero temperature. Suggesting that $T_c$ is much higher than room temperature, the measured at 297 K $2\Delta/e \approx 0.6$ V should be close to $2\Delta(0)$. Then, from (1) one obtains $T_c$ of about 2000 K, which is in good agreement with the model of Little [24] and points out to quasi-1D character of superconductivity. Later in the paper, $T_c$ will be estimated more accurately, taking into account temperature dependence of the energy gap. Although exact mechanism of superconductivity in the case of neural network is not known, equation (1) still should be valid, at least to some approximation, as it comes from the competition between superconducting condensation and thermal energies.

The quasi-1D nanostructures in neurons, in which one could expect superconductivity, are microtubules. It is interesting that they were already suggested as structures responsible for quantum processing of information [5,33,34]. In the nerve cells, microtubules are packed together with shorter neurofilaments [35]. The latter may provide electron-electron interaction necessary to form superconducting state, similar to what is described in the model of Little.

Another suggestion is that superconductivity is the property of water inside the microtubules. The microtubules are hollow tubes with the inner diameter of 14 nm. It is currently not known in what state water is present there. According to a recent analysis of the possibility of room-temperature superconductivity [19], the highest probability to find it is in hydrates. Indeed, the current record-high $T_c$ belongs to $H_3S$ [7]. The presence of hydrogen and other light elements seems to be essential for high $T_c$, but the sulfur in $H_3S$ is not light element. In the periodical table of elements, just above it is similar but lighter element - oxygen. Then, in a certain state, perhaps confined to nano-tubes, $H_2O$, or rather $H_nO$ with n larger than 2 might be superconducting with $T_c$ higher than in $H_3S$.

The importance of water was revealed earlier in experiments performed in [29]. Due to its evaporation, the IV curves plotted in Fig. 2 lasted less than one hour, after which the sample became semiconducting. The underlying experimental reason for such a drawback was small thickness of brain slice (40 micrometers), which, on the other hand, was necessary for formation of local thermo-magnetic instabilities with characteristic double-gap jumps in IV curves [31]. In order to extend the duration of superconducting state, even for the price of losing characteristic jumps, the brain slices of *Sus domesticus* were cut with much higher thickness than rat's brain slices, about 1 mm. Being pressed to current and potential leads, the thickness was, however, somewhat reduced.

The short, one minute, exposure to graphene used in [29], however, did not gave reproducible results for fresh non-preserved *Sus domesticus'* brain slices of this thickness. The IV curves became reproducible only after three hours of exposure to the graphene solution. A set of such curves, recorded one after another with a period of about 2 minutes, for a sample from the outer part of *Sus domesticus'* brain, is shown in the inset of Fig. 5. In spite of some hysteresis, an increase in conductance is seen in most of the curves, and it is centered at about 0.6 V, as marked by the large tilted red arrows.

To better reveal this anomaly in conductance, IV curves in the inset were approximated by polynomials of 9[th] order and differentiated. Several derivatives at both positive and negative voltages are shown in the main plot of Fig. 5 as black lines. It is seen that the main peak in the differential conductance for the majority of the curves (marked with black vertical arrows) is centered at about 0.6 V. If one assumes that the type of connection to the potential electrodes in this experiment is as in Fig. 3c, *i.e.* with suppressed superconductivity close to both of them, then the maximum conductance should be at the voltage $2\Delta/e$ of $\approx 0.6$ V. This is in good agreement with the double-gap voltage jumps found in experiments in Fig. 4.

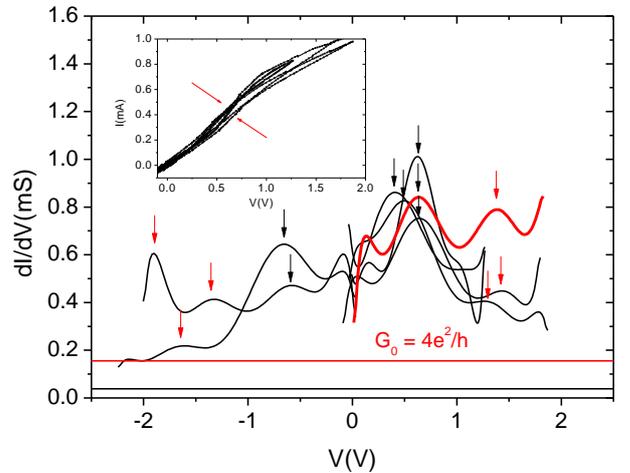

Fig. 5. Differential conductance as function of voltage for a 1-mm thick slice of *Sus domesticus'* brain. The main peak in conductance centered at double-gap voltage $2\Delta/e$ is marked by black arrows. Red arrows mark multiple gap peaks. The red curve shows differential conductance for another sample. The straight red line at the bottom shows the value of the superconducting conductance quantum. Conductance should be above this value to reveal superconductivity as decrease of resistance in the system. The inset shows IV curves at positive voltage corresponding to the black curves of conductance in the main plot.

In addition to the main peak, derivatives reveal other peaks, which are integer multiples of $\Delta/e$ and likely to reflect the appearance of the regions of suppressed superconductivity inside the central area labeled S in Fig. 3c.

These peaks are of smaller amplitude than the main peak and are marked with vertical red arrows.

The small unmarked peak close to V = 0 seen on most of the curves could be the feature of graphene itself or evidence of two-gap superconductivity common to hydrates [8]. The red curve in Fig. 5 belongs to a different sample from the inner part of *Sus domesticus'* brain, which was fixed in formaldehyde and treated with graphene solution for 1 min. Although the brain samples were cut from different regions and received very different treatment (fresh *vs.* fixed), derivatives are very similar, with nearly identical positions of peaks.

The straight red line at the bottom of Fig. 5 corresponds to superconducting conductance quantum $S_0 = 4e^2/h$, where h is the Planck`s constant. In [36], it was shown that for one-dimensional nano-channels cut from the 2D layers, superconductivity is only revealed through the decrease in resistance for conductance values above $S_0$. Most of the lines in Fig. 5 are well above $S_0$.

The shape of the red curve in Fig. 5 shows that treatment in formaldehyde preserves superconductor-like gap features of the sample. More than that, after the treatment, the exposure to graphene solution could be reduced to just one minute. Derivatives of IV curves for yet another sample, fixed in formaldehyde, after one-minute exposure to graphene solution, are shown in Fig. 6. The evidence of the peak at 2Δ/e is overwhelming (marked by black arrow), except for the very first recorded curve (red points), the main peak of which is centered at Δ/e.

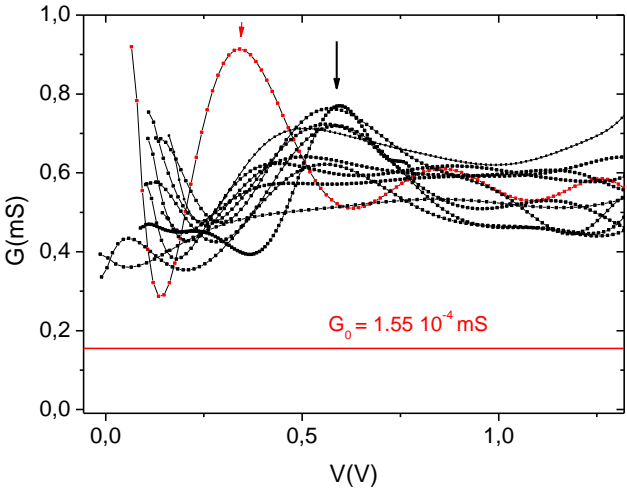

Fig. 6. Differential conductance as function of voltage for the fixed in pharmaldegyde slice of *Sus domesticus'* brain exposed to graphene solution for one minute. Most of the peaks are centered at 2Δ/e (marked by black arrow) except for the very first recorded curve (red points), main peak of which is centered at Δ/e. The straight red line at the bottom of the plot is positioned at the level of the superconducting conductance quantum.

The latter corresponds to the type of connection shown in Fig. 3b. Evidently, during the first measurement, connections of superconducting channels to one of the potential electrodes were without suppression. However, they quickly deteriorated, making more-common connection of the type shown in Fig. 3c. In Fig. 6, it is also noticeable that conductivity tends to increase towards V = 0 for the majority of the curves, which increases the probability of observing dissipation-free part in IV curves at low voltages.

To improve connections of conducting channels to electrodes, the exposure to graphene was combined with a 30-minute ultrasound treatment at the end of the 3-hour graphene exposure of the sample. The result of such treatment is shown in the inset in Fig. 7. The main peak becomes very stable at Δ/e. After many measurements, as it is shown in the main plot of Fig. 7, the 2Δ/e peak appears (black curve). Its amplitude later becomes dominant (red and green curves), and the integer peaks of Δ/e appear too (green and blue curves). Eventually, conductance approaches $G_0$, becoming nearly featureless and voltage-independent (corresponding to the Ohm's law).

The results presented above reveal that using graphene, it is possible to observe superconducting gap corresponding to extremely high $T_c$ of about 2000 K. More accurate value of $T_c$ could be found from (1) taking into account temperature dependence of energy gap. It is also important to make statistical treatment of experimental data starting with double-gap voltage defined as position of peaks in S(V) in Figs. 5-7.

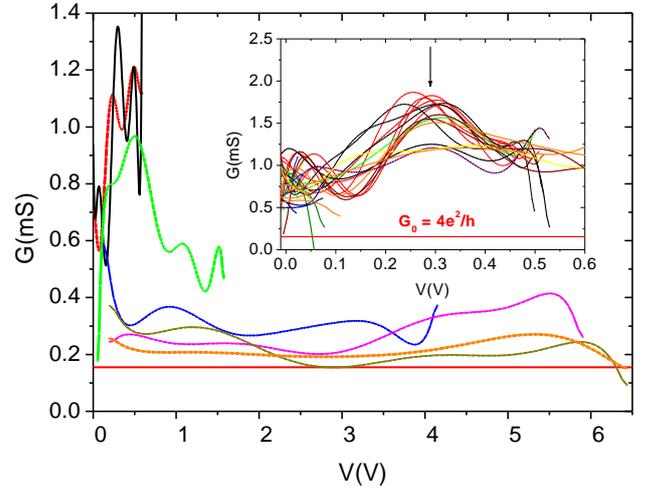

Fig. 7. Differential conductance as a function of voltage for a 1-mm thick slice of *Sus domesticus'* brain treated with ultrasound during the exposure to graphene. All peaks for initial curves are centered at Δ/e, as it is marked by the black arrow in the inset. After a certain time, the 2Δ/e peak appears (black curve in the main plot). Its amplitude becomes dominant (red and green curves), and the integer peaks of Δ/e appear too (green and blue curves). Eventually, conductance approaches the superconducting conductance quantum, which results in nearly featureless and voltage-independent curves.

The mean value of the double-gap voltage for each experiment with its number on the x-axis corresponding to the number of the figure, on which experiment is presented in this paper, together with the standard error in mean is shown in Fig. 8. In experiment 7 (Fig.7), only the curves shown in the inset are analyzed. The single-gap voltage values are also taken into account being multiplied by coefficient of 2.

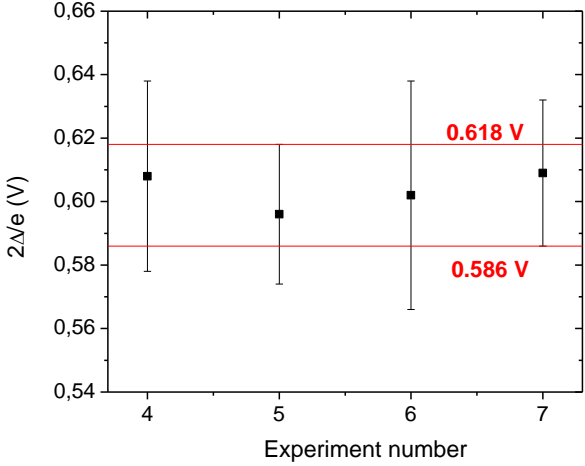

Fig. 8. Mean value of the double-gap voltage together with standard error in mean for experiments in Figs. 4-7 with the number on the x-axis corresponding to the number of the figure, in which corresponding experiment is presented in this paper.

From the Fig.8, one can see that standard error in mean for double-gap voltage is not large, and there is a band ranging from 0.586 to 0.618 V, in which all error bars are overlapped. Since the experiments are of different nature, with results coming from the thermo-magnetic instabilities (experiment 4) and conductance anomalies on NS boundaries (experiments 5-7), and since they are made on brain slices of different mammals and on samples treated in different way, it gives confidence that the observed behaviour is universal and corresponds to real phenomenon that has deep roots in functioning of nervous system.

To estimate critical temperature of superconductor, the mean value of the double-gap voltage, averaged for all four experiments shown in Fig. 8, with the largest error bars encompassing all smaller error bars, are taken for further calculations. These values, which were measured at temperature of 297 K, were fitted to the formula that expresses temperature dependence of the gap [32]:

$$\Delta(T) = \Delta(0)*(1-(T/T_c)^2)^2, \quad (2)$$

combining it with the formula (1) for $\Delta(0)$. The result of this fit is shown in Fig. 9.

The treatment in Fig. 9 gives $T_c$ of 2063 ± 114 K. This value is the main result of the review. If superconductivity in the brain does exist, it is likely to have this critical temperature. The value could be somewhat exaggerated, because in the calculations the possible contribution to gap value from the normal resistance in the areas N shown in Fig. 3 was neglected. Still, it is surprisingly close to the value predicted for the quasi-1D superconductivity in the model of Little [24].

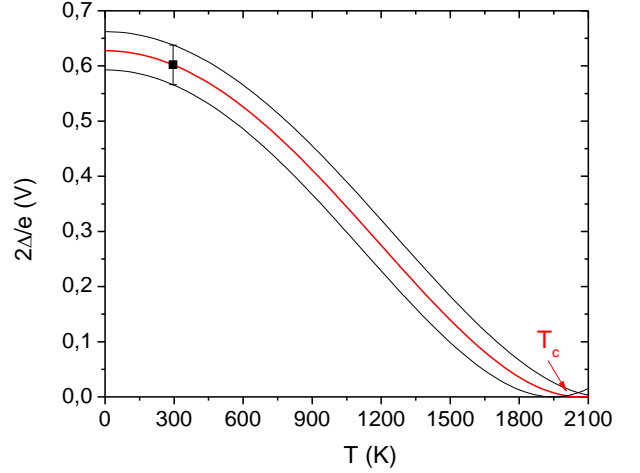

Fig. 9. Measured value of the double-gap voltage fitted to the temperature dependence of the gap given by equation (2) to find mean value and standard error in mean for the critical temperature.

## V. Superconductor-like current-voltage characteristics in brain slices

Whereas extracting gap and estimating $T_c$ was relatively simple, obtaining genuine superconductor-like IV curves for thick slices of pig's brains appeared to be more difficult than for the rat's brain slice, primarily because difficulties of incorporating graphene nanoflakes into the samples. Such difficulties would be expected, because if the superconductivity in the network enables quantum processing of information, it should be well protected from the environment and the superconductivity could be easily destroyed when this protection is lifted, *i.e.* in the attempt to connect electrodes to the network. Still, genuine superconductor-like IV curves were seen for a short time in in the rat's brain in [29], and this gives hope to observe them in the pig's brain slices too.

To address this challenge, different (both fresh and fixed) slices of pig's brains with different exposure to graphene were measured with different materials for electrodes and different distances between them. From the 26 measured samples, superconductor-like IV curves have been observed in 5 samples. In one of the samples, due to its large thickness in comparison with the rat's brain slice measured in [29], such curves lasted for 7 days. Some of these IV curves are shown in Fig. 10 together with the information about the days (in year 2018) on which they were recorded.

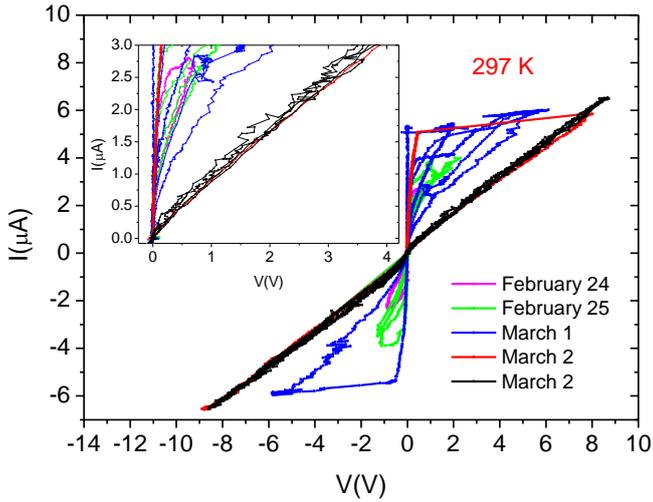

Fig. 10. IV curves of the *Sus domesticus'* brain sample showing superconductor-like features. Information about the dates, on which curves were recorded, is presented in the legend. On the 7$^{th}$ day, too large a current was passed (red curve) and the IV curves became nearly linear (black). In the inset, the low-current parts of the curves are shown to reveal their low-dissipation behaviour.

In the inset, the low-current parts of IV curves are shown to demonstrate their low-dissipation behavior. One can see from the main plot that the critical current density is slightly improving with time, until, on the seventh day, too large a current was passed (red curve) and the IV curves lost their superconducting features (black).

In the measurements above, the influence of a magnetic field of few hundreds of mT on the IV curves was also tested. It was found that a sudden increase in magnetic field frequently lead to an increase of resistance and produced jumps between the resistive states, as shown, for example, by the blue curves in the main plot of Fig. 10 at positive voltages. The amplitude of the voltage jumps was typically bigger than $2\Delta/e$. This behaviour, again, is in agreement with superconductivity in the sample, but also indicates that the system is sufficiently-well protected from the influence of magnetic field.

## VI. Control experiments

To finalize the description of the results, it is important to compare the recorded IV curves with those in the control samples. One of the control samples was a piece of filter paper soaked in graphene solution. It was measured in the same way as the brain slices. The results of its measurements are shown in Fig. 11.

The main plot of Fig. 11 shows voltage dependence of the conductance, and the inset shows IV curves of the sample. IV curves are nonlinear, with a significant increase in resistance at high voltages. The resistance also strongly increases with time, eventually reaching the red line representing superconducting conductance quantum. The $2\Delta/e$ or $\Delta/e$ anomalies are not seen in the plot. There is, however, a systematic kink-like anomaly appearing when voltage is lowered down. There is also a strong increase of conductance at low voltages, which, overlapping with the IV curves of biological samples, may produce the low-voltage peak found in the brain curves. During the degradation, the conductance of the sample quickly drops to $G_0$, but the conductance at low voltages on most curves is high enough to provide a tool for measuring biological samples.

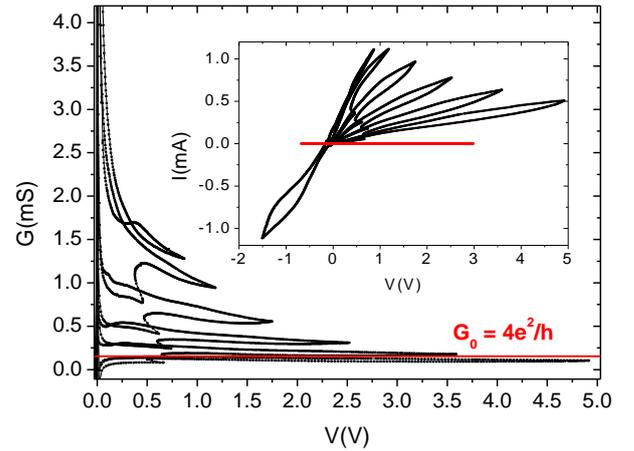

Fig. 11. Conductance as function of voltage for the control sample, filter paper soaked in the graphene solution, measured in the same way as the slices of brain. Inset shows IV curves corresponding to the conductance curves of the main plot. The red line shows the value of the superconducting conductance quantum.

The voltage dependence of conductance for another control sample, a brain slice without graphene, is shown in Fig. 12. Although some wide peaks in conductance as function of voltage are observed, the conductance is too close to the superconducting conductance quantum, and it is difficult to extract any useful information from the plot. Thus, graphene is essential for the experiments.

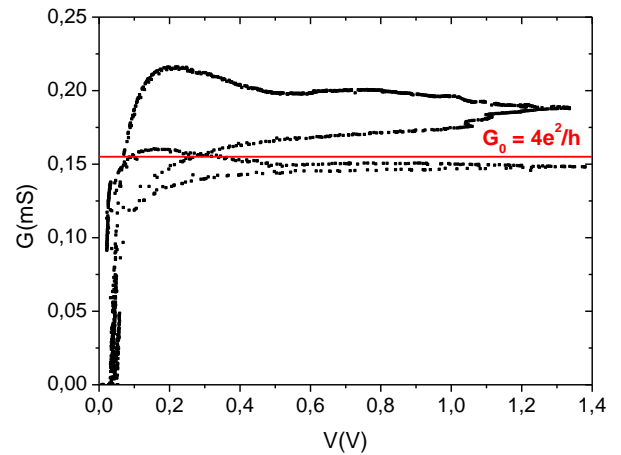

Fig. 12. Conductance as function of voltage for two consecutive increases and decreases of V for a brain slice measured without graphene. The conductance is close to its superconducting quantum value shown by the red line.

## VII. Possible origin of superconductivity in brain

Overall, experiments show that superconductivity with very high $T_c$ is likely to be present in the neural network of the brain. Although it is well protected, it still could be measured using graphene as quantum mediator.

There are indications that water could play essential role in superconductivity. For example, it was shown previously that superconductivity could appear in 3D form at high pressure when ice is doped with nitrogen or other elements [37]. The possibility of superconductivity in confined quasi-1D channels with water was not analyzed.

In the very tightly confined space inside microtubules, water is likely to be structured in a specific way or be in the form of entangled H-O chains, which are schematically shown in Fig. 13. Such chains could have dangling electron bonds linked to well-ordered tubulin proteins [33]. This could provide strong electron-electron interaction necessary for superconductivity with very high $T_c$, like it is in the model of Little [24]. The electron-electron interaction may also come from neurofilaments as was described above.

Alternative explanation is that in the confined space of microtubules, each atom of oxygen is surrounded by several atoms of hydrogen. It was already demonstrated that in hydrates, $T_c$ is dramatically increasing with hydrogen coordination number [9]. The confinement into nano-channels is the main difference between brain structures and already studied hydrates. Such confinement may lead to extremely high critical temperature.

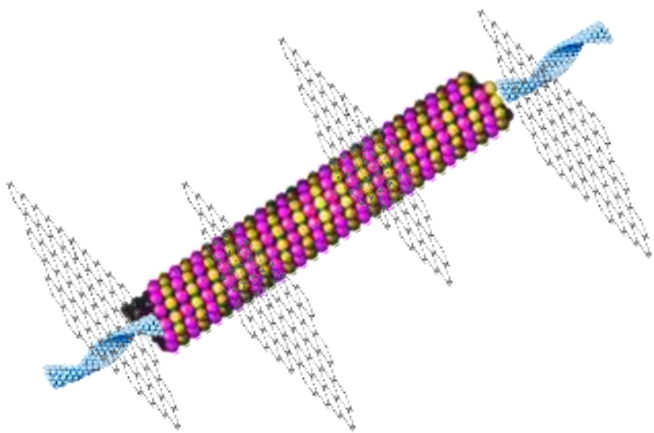

Fig. 13. Schematic representation of a microtubule with a water channel (blue), which could be responsible for superconductivity, and several graphene nanoflakes providing electrical connection to the channel.

The properties of individual microtubules with and without water were measured in [38,39]. It was shown that water is responsible for high conductivity of microtubules, and their coherent electrical behaviour was also demonstrated. These measurements, however, are very complicated and strongly influenced by the measuring circuit. The approach in this paper is very different. Firstly, a very large number of microtubules is measured simultaneously, which greatly reduces measurement error. Secondly, in these measurements, the structure of neural connections is closer to that operating in the living brains. Finally, and most important, graphene is used to extract possible quantum features of the neural network.

While the search for the room-temperature superconductivity continues, nature may have discovered it a long time ago in water, one of the most abundant substances in the universe, and it might be in use to enable, through the enhanced coherence, quantum processing of information. In this sense, intelligent life might be literally dependent on water.

Since it is evident [7,19,37] that high pressure is beneficial for superconductivity, one can speculate that initial structures leading to intelligent life appeared first in the depths of oceans, where the right combination of high pressure, ambient temperature and abundance of water is present. If the mechanism of superconductivity is linked to high coordination number of hydrogen atoms, an extra oxygen from water could be released, which might explain initial high concentration of oxygen on early stages of the evolution of life on the planet.

One could even claim that the nano-scale jigsaw puzzle of intelligent life might be close to completion with realizing that the hydrates-based superconductivity in microtubules could be responsible for quantum processing of information. With superconductivity, quantum processing could follow one of already suggested mechanisms, for example, that described by S. Hameroff and R. Penrose [5,33,34], giving the brain unprecedented computational power.

## Summary

In summary, there is immense progress in search for superconductors with high critical temperature. In particular, graphene-mediated experiments could be treated as providing evidence for superconductivity in brain with extremely high critical temperature of 2063 ± 114 K. Such a high $T_c$ would enable quantum processing of information at room temperature if the temperature of biological structures is kept stable within a modest temperature interval, as it takes place in natural brains. As a coherent quantum phenomenon, superconductivity could be responsible for coherent behaviour of living organisms and even consciousness [1].

The implications of the findings described in the review could be very diverse. Superconductivity, for example, may provide explanation for long-term memory. As the quantum coherent phenomenon, it could form basis of the coherent behavior of living organisms. Finally, the coherent features of the brain and nervous system could be analyzed using well developed and constantly improving models developed for superconductivity.

**Acknowledgments**

Author thanks Prof. M. Fyhn, D. O. Ø. Mjærum and Dr. I. Mikheenko for providing samples for measurements. Dr. Y. Mikheenko is acknowledged for critically reading the paper, and D. O. Ø. Mjærum for useful discussions and help with experiments.